\documentclass[epj]{svjour1}
\usepackage{graphics}
\usepackage{makeidx}   
\usepackage{subeqnar}  

\usepackage{epsfig}

\def\etal{\mbox{{\it et al.}}}

\newcommand{\epem}{\ensuremath{\rm e^+ e^-}}

\newcommand{\Ltre}       {\mbox{L{\sc 3}}}

\newcommand{\Wtolnu}     {\mbox{$\rm{W}\rightarrow
                                 \ell\overline{\nu}_{\ell}$}}

\def\F{{\cal F}}
\def\DRV{\Delta_{\mbox{\tiny R}}^{\mbox{\tiny V}}}

\begin{document}
\title{Quark Mixing, CKM Unitarity}
\subtitle{The unitarity problem}
\author{H. Abele\inst{1} \and E. Barberio\inst{2}\and D. Dubbers\inst{1}\and F. Gl\"uck\inst{3}\thanks{\emph{Present address:}  Johannes Gutenberg University Mainz, Inst. Physics, WA EXAKT,
D-55099 Mainz, Germany }\and J.C. Hardy\inst{4}\and W.J.
Marciano\inst{5}\and A. Serebrov\inst{6}\thanks{\emph{also:} Paul
Scherrer Institut CH-5232 Villigen PSI Switzerland
}\and N. Severijns\inst{7}
%
}                     
\offprints{H. Abele, abele@physi.uni-heidelberg.de}          
\institute{Physikalisches Institut der Universit\"at Heidelberg,
Philosophenweg 12, D-69120 Heidelberg, Germany \and Physics
Department, Southern Methodist University, Dallas, TX, USA \and
Research Institute for Nuclear and Particle Physics, H-1525
Budapest, POB 49, Hungary \and Cyclotron Institute, Texas A\&M
University, College Station, TX 77843, USA\and Brookhaven National
Laboratory Upton, New York 11973, USA \and Petersburg Nuclear
Physics Institute, Gatchina 188350 Leningrad District, Russia \and
Instituut voor Kern- en Stralingsfysica, Katholieke Universiteit
Leuven, Celestijnenlaan 200 D, B-3001 Leuven, Belgium }
\date{}
%
\abstract{In the Standard Model of elementary particles,
quark-mixing is expressed in terms of a 3 x 3 unitary matrix $V$,
the so called Cabibbo-Kobayashi-Maskawa (CKM) matrix. Significant
unitarity checks are so far possible for the first row of this
matrix. This article reviews the experimental and theoretical
information on these matrix elements. On the experimental side, we
find a 2.2 $\sigma$ to 2.7 $\sigma$ deviation from unitarity,
which conflicts with the Standard Model.
} 
\maketitle
\section{Introduction}
\label{intro} The fundamental constituents of matter are the
quarks and the leptons. The quark-mixing Cabibbo-Kobayashi-Maskawa
(CKM) matrix parametrizes the weak charged current interactions of
quarks. The Standard Model does not predict the content of the CKM
matrix and the values of individual matrix elements are determined
from weak decays of the relevant quarks. In this context, weak
decays of nuclei, hadrons and CP violating processes play an
important role. The CKM matrix is required to be unitary. One
possible direct precision test of unitarity involves the top row
of ${\it V}$, namely
\begin{equation} |V_{ud}|^2 + |V_{us}|^2 + |V_{ub}|^2 = 1-\Delta.
\end{equation} In the Standard Model with a unitary CKM matrix, $\Delta$ is zero. The test fails for unknown reasons and a deviation
from unitarity has been found with nuclear
$\beta$-decay~\cite{Hardy} and neutron-$\beta$-decay
data~\cite{Abele02}. Four parameters describe a unitary 3x3
matrix. The "standard" parametrization utilizes 3 angles and one
phase. Table 1 shows the matrix elements of the CKM matrix as
assessed by the Particle Data Group \cite{Hagiwara}. The range
corresponds to 90\% CL limits on the angles and the phase. The
unitarity constraint has pushed $|V_{ud}|$ about two to three
standard deviations higher than given by the experiments.

This article summarizes current knowledge on the CKM matrix from a
workshop at Heidelberg in September 2002 \cite{HADM}. The workshop
reviewed the information to date on the inputs for the unitarity
check from the experimental and theoretical side. Dedicated to
quark-mixing of the first row of the CKM matrix, the meeting
collected complementary information to the CKM workshops held at
CERN and Durham \cite{Battaglia,Durham} and had its main emphasis
on $|V_{ud}|$. Due to its large size, a determination of
$|V_{ud}|$ is most important. It has been derived from a series of
experiments on superallowed nuclear $\beta$-decay through
determination of phase space and measurements of partial lifetimes
as will be explained in section 2. With the inclusion of nuclear
structure effect corrections, a value of $|V_{ud}|$ =
0.9740(5)~\cite{Hardy} emerges in good agreement of different,
independent measurements in nine nuclei. Combined with $|V_{us}|$
= 0.2196(23) from kaon-decays and $|V_{ub}|$ = 0.0036(9) from
B-decays, this leads to $\Delta$ = 0.0031(14), signaling a
deviation from the Unitarity condition by 2.2 $\sigma$ standard
deviations. The quoted uncertainty in $|V_{ud}|$ is dominated by
the uncertainties in the theoretical correction terms, and, as
described in section 2, current nuclear experiments are focused on
testing and refining those correction terms that depend on nuclear
structure. Such terms are avoided entirely in neutron
$\beta$-decay (see section 3) and in pion $\beta$-decay.
\begin{table}
\caption{CKM quark-mixing matrix with 90\%. C.L. from a global fit
to angles and phase~\cite{Hagiwara}. The unitarity constraint has
pushed $|V_{ud}|$ about two to three standard deviations higher
than given by the experiments. }\
\begin{center}
\renewcommand{\arraystretch}{1.4}
\setlength\tabcolsep{5pt}
\begin{tabular}{|l|l|l|}
\hline 0.9741 to 0.9756&  0.219 to 0.226& 0.0025 to 0.0048\\\hline
0.219 to 0.226& 0.9732 to 0.9748& 0.038 to 0.044\\\hline
0.004 to 0.014& 0.037 to 0.044& 0.9990 to 0.9993 \\
\hline

\end{tabular}
\end{center}
\label{apptab1b}
\end{table}
Recently, the mixing of the down quark has been studied in the
decay of free neutrons. With the measurement of the neutron decay
$\beta$-asymmetry $A_0$, using a highly polarized cold neutron
beam with an improved instrument and the world average of the
neutron lifetime $\tau$ one is now capable of extracting a value
for the first entry of the CKM-quark-mixing matrix, whilst
avoiding large corrections to the raw-data or problems linked to
nuclear structure. With neutron-decay data, $|V_{ud}|$ =
0.9717(13) leads to significant deviation $\Delta$ = 0.0076(28),
which is 2.7 times the stated error. The pion $\beta$-decay has
been measured recently at the PSI. The pion has a different hadron
structure compared with neutron or nucleons and it offers another
possibility in determining $|V_{ud}|$. The preliminary result is
$|V_{ud}|$=0.9771(56)~\cite{Pocanic}. The error is still too large
to allow a significant unitarity check.

A violation of unitarity in the first row of the CKM matrix is a
challenge to the three generation Standard Model. The data
available so far do not preclude there being more than three
generations; CKM matrix entries deduced from unitarity might be
altered when the CKM matrix is expanded to accommodate more
generations \cite{Hagiwara,Marciano1}. A deviation $\Delta$ has
been related to concepts beyond the Standard Model, such as
couplings to exotic fermions \cite{Langacker1,Maalampi}, to the
existence of an additional Z boson \cite{Langacker2,Marciano2}, to
supersymmetry or to the existence of right-handed currents in the
weak interaction \cite{Deutsch}. Non-unitarity of the CKM matrix
in models with an extended quark sector give rise to an induced
neutron electric dipole moment that can be within reach of the
next generation of experiments \cite{Liao}.

\section{$|V_{ud}|$ from superallowed $0^{+} \rightarrow 0^{+}$ beta decay}

Currently, superallowed $0^{+} \rightarrow 0^{+}$ nuclear
$\beta$-decay provides the most precisely determined value for
$V_{ud}$, the up-down quark mixing element of the
Cabibbo-Kobayashi-Maskawa (CKM) matrix.  This value is also the
most precise result for any element in the CKM matrix and leads to
the most demanding test available of CKM unitarity, a test which
apparently fails by more than two standard deviations
\cite{TH98,TH02}: With $|V_{us}|$ = 0.2196(23) and the negligibly
small $|V_{ub}|$ = 0.0036(9), one gets

\begin{equation}
 |V_{ud}|^2 + |V_{us}|^2 + |V_{ub}|^2 = 1 - \Delta = 0.9969 \pm
 0.0014,
\label{Un}
\end{equation}

\noindent and $\Delta$ = 0.0031(14). Nuclei have the singular
advantage that transitions with specific desirable characteristics
can be selected and then isolated for study. The case of $0^{+}
\rightarrow 0^{+}$ $\beta$-transitions between $T=1$ analog states
is an excellent example, since angular momentum limits such
transitions uniquely to the vector part of the weak interaction,
thus eliminating the need for special experiments designed to
account separately for the vector and axial-vector parts.
Furthermore, the nuclear matrix element is given by the
expectation value of the isospin ladder operator and, since the
parent and daughter states for these transitions are analogs of
one another, the result should be independent of any details of
nuclear structure to the extent that isospin symmetry is
preserved.  Of course, there are corrections to this simple
picture, originating from charge-dependent mixing and other
electromagnetic effects, but these corrections are small -- of
order $1\%$ -- and calculable.

The measured intensity of a particular $\beta$-transition is
expressed as an $ft$-value.  This $ft$-value is determined by
three measured parameters: the transition energy $Q_{EC}$, which
is used in calculating the statistical rate function, $f$; the
half-life of the $\beta$-emitter and the branching ratio for the
transition of interest, which together yield the partial
half-life, $t$.  The experimentally determined $ft$-value relates
to the Fermi constant, $G_F$ via the relationship \cite{TH98,TH02}

\begin{equation}
\F t \equiv ft (1 + \delta_R^{\prime} + \delta_{NS})(1 - \delta_C
) = \frac{K}{2 |V_{ud}|^2G_F^2 (1 + \DRV )} , \label{Ft}
\end{equation}
\noindent where $K$ is a known constant. The small correction
terms comprise $\delta_C$, the isospin-symmetry-breaking
correction; $\delta_R^{\prime}$ and $\delta_{NS}$, the
transition-dependent parts of the radiative correction; and $\DRV
$, the transition-independent part.  Here we have also defined $\F
t$ as the ``corrected" $ft$-value.  Note that, of the four
calculated correction terms, two -- $\delta_C$ and $\delta_{NS}$
-- depend on nuclear structure and their influence in
Eq.(\ref{Ft}) is effectively in the form ($\delta_C -
\delta_{NS}$).

To date, there are nine nuclear $0^{+} \rightarrow 0^{+}$
transitions whose $ft$-values have been measured with a precision
of $\sim 0.1\%$ or better.  Many separate measurements and
experimental teams have contributed to this body of data and the
results can be considered extremely robust, most input data having
been obtained from several independent and consistent measurements
\cite{TH98,Ha90}.  The decay parents -- $^{10}$C, $^{14}$O,
$^{26m}$Al, $^{34}$Cl, $^{38m}$K, $^{42}$Sc, $^{46}$V, $^{50}$Mn
and $^{54}$Co -- also span a wide range of nuclear masses.
Nevertheless, as anticipated by the Conserved Vector Current
hypothesis, CVC, all nine yield consistent $\F t$-values and hence
a unique value for $G_V$.

The value of $V_{ud}$ is obtained by relating the vector constant,
$G_V$, determined from the self-consistent nuclear $\F t$-values,
to the weak coupling constant from muon decay.  This result,
$V_{ud} = 0.9740 \pm 0.0005$, leads to the unitarity test already
displayed in Eq. (\ref{Un}).  (In deriving these results we have
used the Particle Data Group's \cite{Hagiwara} recommended values
for the muon coupling constant and for $V_{us}$ and $V_{ub}$.)  It
is informative to dissect the contributions to the uncertainty
obtained for $V_{ud}$.  The contributions to the overall $\pm
0.0005$ uncertainty are 0.0001 from experiment, 0.0001 from
$\delta_R^{\prime}$, 0.0003 from ($\delta_C  - \delta_{NS}$), and
0.0004 from $\DRV$.  Thus, if the unitarity test is to be
sharpened, then the most pressing objective must be to reduce the
uncertainties on $\DRV$ and ($\delta_C  -  \delta_{NS}$).

Improvements in $\DRV$ are a purely theoretical challenge, the
solution of which will not depend on further experiments.
However, experiments can play a role in improving the next most
important contributor to the uncertainty on $V_{ud}$, namely
($\delta_C  -  \delta_{NS}$).  Clearly this correction applies
only to the results from superallowed beta decay and, in the event
that improvements are made in $\DRV$, will then limit the
precision with which $V_{ud}$ can be determined by this route.
Recently, a new set of consistent calculations for ($\delta_C  -
\delta_{NS}$) have appeared \cite{TH02} not only for the nine well
known superallowed transitions but for eleven other superallowed
transitions that are potentially accessible to precise
measurements in the future.  Experimental activity is now focused
on probing these nuclear-structure-dependent corrections with a
view to reducing the uncertainty that they introduce into the
unitarity test.

\subsection{The future}
The approach being taken by current experiments is to
choose as yet unmeasured superallowed transitions for which it is
predicted that the structure-dependent corrections, ($\delta_C  -
\delta_{NS}$), are particularly large, or to choose several such
transitions that cover a wide range of calculated corrections.  If
a transition with a much larger correction than any currently
known yields an $\F t$-value that also agrees with the current
average -- {\it i.e.} is consistent with CVC -- then this would
constitute a critical test of the accuracy of the calculated
structure-dependent corrections. If such measurements are found to
support the calculations, then this would validate those
calculations and act to reduce the uncertainties attributed to
them, uncertainties that currently are based only on theoretical
estimates.

Experimental attention is currently focused on two series of
$0^{+}$ nuclei: the even-$Z$, $T_z = -1$ nuclei with $18 \leq A
\leq 42$, and the odd-$Z$, $T_z = 0$ nuclei with $A \geq 62$.
Both regions include transitions with larger calculated values
\cite{TH02} for ($\delta_C - \delta_{NS}$) than any of the nine
currently well-known transitions. Of the heavier $T_z = 0$ nuclei,
$^{62}$Ga and $^{74}$Rb are receiving the greatest attention at
this time (see ref. \cite{HT02} and experimental references
therein).  The decays of nuclei in this series are of higher
energy than any previously studied and each therefore involves
numerous weak Gamow-Teller transitions in addition to the
superallowed transition \cite{HT02}.  Branching-ratio measurements
are thus very demanding, particularly with the limited intensities
likely to be available initially for most of the rather exotic
nuclei in this series.  Nevertheless, with the help of shell-model
calculations \cite{HT02}, a combination of detailed $\beta$- and
$\gamma$-ray spectroscopic measurements has recently been used to
obtain a precise value for the superallowed branching ratio in the
decay of prolifically produced $^{74}$Rb \cite{Pi03}.

More accessible in the short term are the $T_z = -1$ superallowed
emitters with $18 \leq A \leq 42$.  The nuclear-model space used
in the calculation of ($\delta_C - \delta_{NS}$) for these nuclei
is exactly the same as that used for some of the nine transitions
already studied.  If the wide range of values predicted for the
corrections are confirmed by the measured $ft$-values, then it
will do much to increase our confidence (and reduce the
uncertainties) in the corrections already being used.   To be
sure, these decays also provide an experimental challenge,
particularly in the measurement of their branching ratios, which
involve {\it strong} Gamow-Teller transitions, but sufficiently
precise results have just been obtained \cite{Ha02} for the half
life and superallowed branching ratio for the decay of $^{22}$Mg
and work on $^{34}$Ar decay is well advanced. New precise
$ft$-values should not be long in appearing.  It would be
virtually impossible for them to have any effect on the central
value already obtained for $V_{ud}$ but they may be expected
ultimately to lead to reduced uncertainties on that value.

\section{$|V_{ud}|$ from neutron $\beta$-decay}

Recently, $|V_{ud}|$ has been derived not from nuclear
$\beta$-decay but from neutron decay data \cite{Abele02}. In this
way, the unitarity check of (1) is based solely on particle data,
i.e. neutron $\beta$-decay, K-decays, and B-decays, where nuclear
structure effects are absent. So much progress has been made using
highly polarized cold neutron beams with an improved detector
setup that one is now capable of extracting a value $V_{ud}$ =
0.9717(13). Here, the unitarity test \begin{equation} |V_{ud}|^2 +
|V_{us}|^2 + |V_{ub}|^2 = 1 - \Delta = 0.9924(28)
\end{equation} fails by $\Delta$ = 0.0076$\pm$0.0028, or 2.7 times the stated
error $\sigma$. Earlier experiments
\cite{Bopp,Yerozolimsky,Schreckenbach} gave significantly lower
values for the neutron decay $\beta$-symmetry $A_0$. Averaging
over the new result and previous results, the Particle Data Group
\cite{Hagiwara} arrives at a new world average for $|V_{ud}|$ from
neutron $\beta$-decay which leads to a 2.2 $\sigma$ deviation from
unitarity.

Since the Fermi decay constant is known from muon decay, the
Standard Model describes neutron $\beta$-decay with only two
additional parameters. One parameter is the first entry $|V_{ud}|$
of the CKM-matrix. The other one is $\lambda$, the ratio of the
vector coupling constant and the axial vector constant. In
principle, the ratio $\lambda$ can be determined from QCD lattice
gauge theory calculation, but the results of the best calculations
vary by up to 30\%. In neutron decay, several observables are
accessible to experiment, which depend on these parameters, so the
problem is overdetermined and, together with other data from
particle and nuclear physics, many tests of the Standard Model
become possible. $|V_{ud}|$ results significantly from the neutron
lifetime $\tau$ and the $\beta$-asymmetry parameter $A_0$. The
$ft$-value is given by
\begin{equation}
ft(1+\delta_R^\prime)=\frac{K}{|V_{ud}|^2G_F^2(1+3\lambda^2)(1+\Delta_R)}
,
\end{equation}
where $f$ =1.6886 is the phase space factor. The model independent
radiative correction $\delta_R^\prime$ and other small terms
changes the phase space factor by 1.5\% to $f^R$ = 1.71335(15)
\cite{Wil82,acwm}. $\Delta_R$ = 0.0240(8) is the model dependent
radiative correction to the neutron decay rate
\cite{Towner1,Hardy}. The $\beta$-asymmetry $A_0$ is a simple
function of $\lambda$, the ratio of the axial vector to vector
coupling constant
\begin{equation}
A_0=-2\frac{\lambda(\lambda+1)}{1+3\lambda^2},
\end{equation} where we have assumed that $\lambda$ is real.

With recent experiments \cite{Abele02,Abele}, one obtains \\$A_0$
= -0.1189(7) and $\lambda$ = -1.2739(19). With this value, and the
world average for $\tau$ = 885.7(7) s, one finds that $|V_{ud}|$ =
0.9717(13). The main contribution to the overall $\pm0.0013$
uncertainty is the experimental error from the $\beta$-asymmetry
$A_0$ with $\pm0.0012$. In these $\beta$-asymmetry experiments,
the total correction to the raw data is 2.0\%. We favour this
result over earlier experiments
\cite{Bopp,Yerozolimsky,Schreckenbach}, where large corrections
had to be made for neutron polarization, electron-magnetic mirror
effects or background, which were all in the 15\% to 30\% range.
The world average value for the neutron mean lifetime includes 11
individual measurements, using different techniques summarized as
"beam methods" and "bottle methods". The measurements agree nicely
with a $\chi^2$ of 0.95 but the world average value is dominated
by a single experiment \cite{Arzumanov}. The error contribution to
$|V_{ud}|$ with $\pm0.0004$ from the neutron lifetime is small and
the same as the error from $\Delta_R$. The contribution to $f$
from $\delta_R^\prime$ is with $\pm0.00004$ completely negligible.

For comparison, information about $|V_{ud}|$ and $\lambda$ are
shown in Fig. (1). The bands represent the one sigma error of the
measurements. The $\beta$-asymmetry $A_0$ in neutron decay depends
only on $\lambda$, while the neutron lifetime $\tau$ depends both
on $\lambda$ and $|V_{ud}|$. The intersection between the bands
derived from $\tau$ and $A_0$ defines $|V_{ud}|$ within one
standard deviation, which is indicated by the error ellipse. For
comparison, $|V_{ud}|$ derived from nuclear $\beta$-decay as well
as a value $|V_{ud}|$ = $\sqrt{1-|V_{us}|^2+|V_{us}|^2}$ predicted
from unitarity are shown, too.

\begin{figure}
\hbox to\hsize{\hss
\includegraphics[width=\hsize]{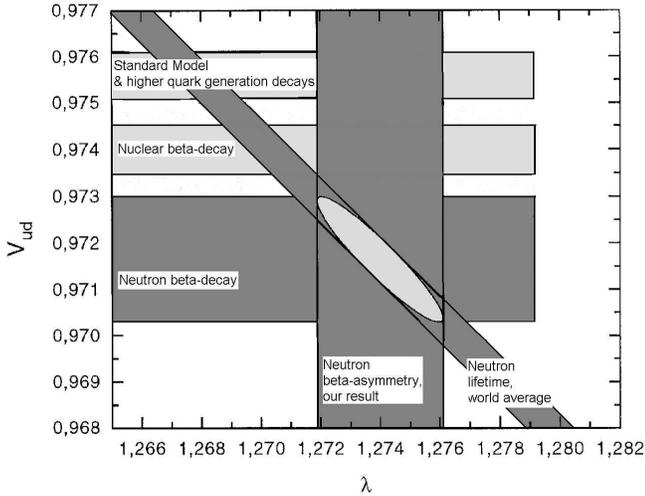}
\hss} \caption{$|V_{ud}|$  vs. $\lambda$. $|V_{ud}|$ was derived
from higher quark generation decays via $|V_{ud}|$ =
$\sqrt{1-|V_{us}|^2+|V_{us}|^2}$ predicted from unitarity, from
$Ft$ values of nuclear $\beta$-decays, and neutron $\beta$-decay.}
\end{figure}

\subsection{The future} The main corrections
in recent neutron decay asymmetry experiments \cite{Abele02} are
due to neutron beam polarization (1.1\%), background (0.5\%) and
flipper efficiency (0.3\%). The total correction is 2.0\%, the
total relative error is 0.68\%. For the future, the plan is to
further reduce all corrections. Major improvements both in neutron
flux and degree of neutron polarization have already been made:
First, the new ballistic supermirror guide at the Institute-Laue
Langevin in Grenoble gives an increase of about a factor of 4 in
the cold neutron flux~\cite{Haese}. Second, a new arrangement of
two supermirror polarizers allows to achieve an unprecedented
degree of neutron polarization $P$ of between 99.5\% and 100\%
over the full cross section of the beam~\cite{Soldner}. Third,
systematic limitations of polarization measurements have been
investigated: The beam polarization can now be measured with a new
method using an opaque $^3$He spin filter with an uncertainty of
0.1\% \cite{Heil,Zimmer}. As a consequence, we will be able to
improve on the main uncertainties in reducing the main correction
of 1.1\% to less than 0.5\% with an error of 0.1\%. Future trends
have been presented on the workshop "Quark-mixing, CKM Unitarity"
\cite{HADM}. Regarding lifetime measurements, several independent
experiments with a projected accuracy of one second or better are
carried out. One of them uses the storage of ultra-cold neutrons
in a material trap with a gravitational valve \cite{Fomin}. The
coating of the trap surface allows to obtain a storage time of
870s, which is very close to the neutron lifetime. The use of two
traps with different sizes and the method of size extrapolation
expect an accuracy for the neutron lifetime at the level of one
second. In a different approach, neutrons are trapped magnetically
with permanent magnets \cite{Ezhov} or with the superconducting
magnets \cite{Hartmann,Huffman}. Regarding the Unitarity problem,
about half a dozen new instruments are planned or are under
construction for beta-neutrino correlation $a$ and
beta-correlation  $A$ measurements at the sub-10$^{-3}$ level,
which should result in a value of $|V_{ud}|$ whose error is
dominated by the theoretical uncertainties in the radiative
corrections (see section 4).

\section{Electroweak radiative corrections}

In order to obtain accurate values for the $V_{ud}$ element of the
CKM matrix from nuclear, neutron and pion beta decays, we need
precise radiative correction calculations for their observables.
For this purpose, one has to deal with several different kinds of
Feynman diagrams. In addition to the $W$ boson mediating the beta
decay process, further electroweak bosons (photon, $W$ and $Z$
bosons) can be created and absorbed by the fermions, and these
bosons can change slightly the various decay probabilities.

In order to compute the radiative corrections, both the virtual
and the photon bremsstrahlung contributions have to be evaluated.
Usually, the bremsstrahlung photons are not detected in beta decay
experiments, therefore one has to integrate these photons down to
zero energy. These bremsstrahlung integrals are infinite: they
have infrared divergency. However, adding the contributions of the
virtual diagrams to the bremsstrahlung integrals, the infrared
divergency disappears. The bremsstrahlung photons can leave the
small space-time region of the beta decay, and their momenta can
be of similar magnitude as the momenta of the other particles
involved in the decay process. Therefore, these photons can change
the beta decay kinematics (in contrast to the virtual photons,
which have no effect to the kinematics). It is important to take
into account this fact, in order to obtain meaningful radiative
correction results. We refer to Refs. \cite{Gluck94,Gluck98b} for
detailed explanation of this photon bremsstrahlung kinematics
effect. The photon bremsstrahlung calculation is theoretically
simple and reliable. Due to their small energy, the bremsstrahlung
photons see only the hadron charges, therefore this part of the
radiative correction calculation has no uncertainties related with
strong interaction models. On the other hand, technically it is
more complicated. In order to evaluate the many-dimensional
integrals occurring in the computation of arbitrary observables,
the Monte Carlo method seems to be expedient to use
\cite{Gluck97a}.

In order to compute the virtual correction, the energy and
momentum of the virtual photon has to be integrated from zero to
infinity. The small energy part of the virtual integrals has
similar properties to the bremsstrahlung correction (infrared
divergence, almost no strong interaction dependence, sensitivity
to external particle momenta). \newline There\-fore, it is
expedient to separate this part of the virtual correction from the
larger photon energy region contributions. Sirlin has introduced
in his
 prominent paper \cite{Sirlin67} a separation of the order-$\alpha$
 radiative correction into model independent (MI) and model dependent (MD)
parts. The MI (outer) correction is defined as the sum of the
photon bremsstrahlung correction and the small energy part of the
virtual correction. The MD (inner) correction contains the medium
and large energy (asymptotic) parts of the virtual contributions.
It was proved in Ref. \cite{Sirlin67} that neglecting some small
terms of order 0.01 \%, the MD correction can be absorbed into the
dominant vector and axial vector form factors and is not taken
into account in a determination of $\lambda$ from the
$\beta$-asymmetry parameter $A_0$.

The order-$\alpha$ MI radiative corrections to the total decay
rates of allowed beta decays can be simply computed by the
universal Sirlin function \cite{Sirlin67}. For example, in neutron
decay this correction increases the decay rate by 1.5 \%. On the
other hand, the MI correction to the asymmetry parameters is
rather small \cite{Shann,Gluck92,Gluck96,Gluck98b}. Due to this
correction, the value of the parameter $\lambda$ from a
measurement of the $\beta$-asymmetry parameter is changed by
0.0003, which is far below the present experimental error of
0.0019. Other observables for determining $\lambda$ have larger
corrections. For example for the proton spectrum in unpolarized
neutron decay, the change of $\lambda$ is 0.01 due to the MI
correction \cite{Gluck93} and the main part of this large
correction comes from the photon bremsstrahlung kinematic effect
mentioned above.

\subsection{One and Two Loop Electroweak Corrections} Modulo the
Fermi function, electroweak radiative corrections to superallowed
$0^{+} \rightarrow 0^{+}$ nuclear beta decays are traditionally
factored into two contributions called inner and outer
corrections.  The outer (or long distance) correction 1 +
$\delta_R^{\prime}$ + $\delta_{NS}$ is given by
\begin{equation}
1~+~\frac{\alpha}{2\pi}(g(E, E_{max})~+ 2C_{NS})~+ \delta_{2}(Z,
E) \label{al2pi},
\end{equation}
where $g(E, E_{max})$ is the universal Sirlin function
\cite{Sirlin67} which depends on the nucleus through $E_{max}$,
the positron or electron end point energy. $C_{NS}$ is a nuclear
structure dependent contribution induced by axial-current
nucleon-nucleon interactions \cite{itjh} and $\delta_{2}$ is an
$O(Z\alpha^{2})$ correction partly induced by factorization of the
Fermi function and outer radiative corrections \cite{wjgr}.

The contribution from $g(E, E_{max})$ is quite large
$(\sim1.3\%~for~O^{14})$ due to a $3\ln(m_{p}/E_{max})$ term which
generally dominates.  Summation of $(\alpha \ln
m_{p}/E_{max})^{n}$, $n = 2,3\dots$ contributions from higher
orders gives an additional $0.028\%$ correction \cite{acwm} while
additional $O(\alpha^{2})$ effects are estimated to be $<0.01\%$.

$C_{NS}$ and $\delta_{2}(Z, E)$ are nucleus dependent.  The
leading contribution to $\delta_{2}$ is of the form
$Z\alpha^{2}\ln m_{p}/E$ where $Z$ is the charge of the daughter
nucleus.  Just as in the case of the Fermi function, $\delta_{2}$
is usually given for positron emitters (since that is appropriate
for superallowed decays).  For electron emitters the sign of $Z$
should be changed in both the Fermi function and $\delta_{2}(Z,
E)$. Unfortunately, as pointed out by Czarnecki, Marciano and
Sirlin \cite{acwm}, that sign change was not made in the case of
neutron decay.  As a result, the often quoted $0.0004$
contribution from $\delta_{2}$ to neutron $\beta$-decay should be
changed to $- 0.00043$, an overall shift of $-0.083\%$. With those
corrections, the overall uncertainty in the outer radiative
corrections is now estimated to be about $\pm0.01\%$.

The inner radiative correction factor 1 + $\DRV$ is given (at one
loop level) by
\begin{equation}
1~+~\frac{\alpha}{2\pi}(4 \ln \frac{m_{Z}}{m_{p}} + \ln
\frac{m_{p}}{m_{A}} + A_{g} + 2C), \label{lnmp}
\end{equation}
where the $\frac{2\alpha}{\pi}\ln m_{Z}/ m_{p} \simeq 0.0213$
universal short- distance correction dominates \cite{rmp}.  The
contributions induced by axial- vector effects are relatively
small but carry the bulk of the theoretical uncertainty
\begin{equation}
\frac{\alpha}{2\pi}[\ln \frac{m_{p}}{m_{A}} + A_{g} + 2C] {\simeq}
{-–}0.0015 \pm 0.0008. \label{magc}
\end{equation}
It stems from an uncertainty in the effective value of $m_{A}$
that should be employed. The quoted uncertainty in eq. (3) allows
for a conservative factor of 2 uncertainty in that quantity.  It
would be difficult to significantly reduce the uncertainty for
nuclei or the neutron.  In the case of pion beta decay, the
uncertainty is likely to a factor of 2 or more smaller.

High order $(\alpha \ln m_{Z}/m_{p})^{n}, n = 2,3\dots$ leading
log contributions are expected to dominate the multi-loop effects.
They have been summed by renormalization group techniques
\cite{Marciano1}, resulting in an increase in eq. (8) by $0.0012$.
Next to leading logs of $O(\alpha^{2}\ln m_{Z}/m_{p})$ have been
estimated to give $-0.0002\pm0.0002$ while $O(\alpha^{2})$ effects
are expected to be negligible.  In total, a recent update finds
\cite{acwm}
\begin{equation}
Inner R. C. Factor = 1.0240\pm0.0008, \label{inrc}
\end{equation}
which is essentially the same as the value given by Sirlin in 1994
\cite{lwsi}. It leads to
\begin{equation}
|V_{ud}|=0.9740\\\pm0.0001\pm0.0001\pm0.0003\pm0.0004,
\label{vud7}
\end{equation}
extracted from super-allowed beta decays, where the errors stem
from the experimental uncertainty, the two transition dependent
parts of the radiative corrections $\delta_R^{\prime}$ and
$\delta_C - \delta_{NS}$, and the inner radiative correction
$\DRV$ respectively.

In the case of neutron decay, the radiative corrections carry a
similar structure and uncertainty.  Correcting for the sign error
in the $Z\alpha^{2}$ effect, one finds the master formula
\cite{acwm}
\begin{equation}
|V_{ud}|^{2}=\frac{4908 \pm 4 sec}{\tau_{n}(1 +3\lambda^2)}.
\label{4908}
\end{equation}
Employing $\tau_{n} = 885.7(7)$s and $\lambda$ = 1.2739(19) then
implies
\begin{equation}
|V_{ud}| = 0.9717\pm0.0004\pm0.0012\pm0.00004\pm0.0004
\label{0971}
\end{equation}
where the errors stem from the experimental uncertainty in the
neutron lifetime, the $\beta$-asymmetry $A_0$ and the theoretical
outer and inner radiative correction $\delta_R^{\prime}$ and
$\DRV$ respectively.  In the case of pion beta decay, the theory
uncertainty in $|V_{ud}|$ is probably $\pm0.0002$ or smaller, but
the small $(\simeq 10^{-8})$ branching ratio makes a precision
measurement very difficult.

\section{$|V_{us}|$ from hyperon and kaon-decays}

Hyperon or K decays determine $V_{us}$. The analysis of hyperon
data has larger theoretical uncertainties because different
calculations of SU(3) symmetry breaking effects disagree.
Therefore the Particle Data Group relies on the K decay data,
based on a derivation in 1984 \cite{Leutwyler}.

The rates of $K_{l3}$ has the form \cite{Leutwyler,Bargiotti}
\begin{equation}
\Gamma=\frac{G_F^2}{192\pi^3}M_k^5|V_{us}|^2C|f(0)|^2|I(1+\delta)(1+\Delta)
\end{equation}
 with phase space integral $I$, form factor $f$, $M_k$ the kaon mass, Clebsch-Gordan coefficient $C^2$ and radiative corrections \cite{Mar86,Williams} $\Delta$ = 2.12$\pm$0.08\% and $\delta$ = -2\% for $K_{e3}^+$ decays and $\delta$ = 0.5\% for $K_{e3}^0$ decays.
Updates of $|V_{us}|$ with revised radiative corrections
\cite{updatevus} are in agreement with the current PDG value
$|V_{us}|$ = 0.2196$\pm$0.0023 \cite{Leutwyler,Hagiwara} and even
indicate a decrease of the central value by up to 1\%. However, a
very recent report of the $K^{+}_{e3}$ branching ratio from E865
at BNL results in a larger $|V_{us}|$ = 0.2272(30) \cite{Sher}.
With this value of $|V_{us}|$ alone, we find no significant
deviation from CKM unitarity. On the other hand, the discrepancy
between this BNL $|V_{us}|$ value and a value from $K^{0}_{e3}$ is
on the 3 $\sigma$ level. New, ongoing or prepared $K_{e3}$
measurements (e.g. CMD2, NA48, KTEV, KLOE) \cite{HADM,Durham} will
help to solve this K-decay problem. After this workshop, a new
analysis of hyperon decays by Cabibbo, Swallow and Winston that is
insensitive to first order SU(3) breaking effect appeared
\cite{Cabibbo}. It found $|V_{us}|$ = 0.2250(27) which is in
better agreement with unitarity and the E865 $K_{e3}$ results than
previous studies, thus providing additional motivation for further
experimental work on $|V_{us}|$.

\section{CKM matrix elements from decays of W bosons and top quarks}
Decays of W$^\pm$ bosons produced at LEP2 have been used to
measure the Cabibbo-Kobayashi-Maskawa matrix element  $|\rm
V_{cs}|$ with a precision of 1.3\% without the need of a form
factor. The same data set has been used to test the unitarity of
the first two rows of the matrix at the 2\% level. At a future
$\rm e^+ e^-$ linear collider, with a data sample of few million
of W decays a precision of 0.1\% can be reached.

\subsection{Determination of the W branching ratio}
%
\renewcommand{\arraystretch}{1.2}
\begin{table}
\begin{center}
\begin{tabular}{|l|l|l|}
\hline
 &{current uncertainty} & projected for new collider \\
\hline
$|V_{ud}|$  & $\pm0.0005$ & $\pm0.0028$   \\
$|V_{us}|$\ & $\pm0.0023$ & $\pm0.0124$   \\
$|V_{ub}|$   & ~$\pm0.008$& $\pm0.011$   \\
$|V_{cd}|$  & $\pm0.016$ & $\pm0.0072$    \\
$|V_{cs}|$  & $\pm0.16$ & $\pm0.0017$    \\
$|V_{cb}|$  & $\pm0.0019$ & $\pm0.11$    \\
$|V_{td}|$  & $|V_{td}|/|V_{ts}|<0.24$ & $\pm0.026 \pm0.35$   \\
$|V_{ts}|$  & $\pm0.008$ & $\pm0.006 \pm0.0002$    \\
$|V_{tb}|$  & $+0.29,-0.12)$ & $\pm0.000008\pm0.005$    \\
\hline
 \end{tabular} \caption{\small
Current and expected precision of CKM matrix elements. Prospective
measurements are listed for linear $e^+e^-$ colliders. The second
column is taken from ~\cite{pm}. The second error is due to an
uncertainty of $\Gamma_{top}$ of 1\%. \label{wwbra}}
\end{center}
\vspace*{-0.5truecm}
\end{table}
The observation of W decays offers another way to determine the
CKM matrix elements. In the Standard Model the branching fraction
of the W boson decays depends on the six CKM matrix elements which
do not involve the top quark. Measuring the W production rates for
different flavours gives access to the individual CKM matrix
elements without
  parameterization of non-perturbative QCD:
$$
  \Gamma(\rm W \to q' \bar{q}) = \frac{C(\alpha_s)G_F M^3_W}{6\cdot \sqrt{2}\pi}
          |V_{ij}|^2 = (707 \pm 1) |V_{ij}|^2 \rm MeV, $$
where
$$
          C(\alpha_s) = 3 \bigg[ 1 +
          \sum_{i=1,3}
          \frac{a_i \alpha_s(\rm{M}^2_{\rm{W}})}{\pi}
          \bigg]
$$
is the QCD colour factor, up to the third order in
 $\alpha_s(\rm{M}^2_{\rm{W}})$, the strong coupling
constant.  Furthermore, 'on shell' W bosons decay before the
hadronization process starts, and the quark transition occurs in a
perturbative QCD regime.
Hence, W boson decays offer a complementary way to determine the
CKM matrix elements.

In a similar way one can relate the top quark transition
$B(t\rightarrow qW)$ to the CKM matrix elements $|V_{tq}|$.

From 1997 to 2000 the LEP \epem collider has been operated at
energies above the threshold for W-pair production. This offered a
unique opportunity to study the hadronic decays of W boson in a
clean environment and to investigate the coupling strength of W
bosons to different quark flavours.

The leptonic branching fraction of the W boson
\\ ~$\mathcal{B}(\Wtolnu)$ is related to the six CKM elements not
involving the top quark by:
$$
  \frac{1}{\mathcal{B}(\Wtolnu)}\quad = \quad
   3\Bigg\{1+\bigg[ 1 + \frac{\alpha_s(\rm{M}^2_{\rm{W}})}{\pi}
          \bigg]
          \sum_{\tiny\begin{array}{c}i=(u,c),\\j=(d,s,b)\\\end{array}}
          |\rm{V}_{ij}|^2
  \Bigg\}.
$$
Using  $\alpha_s(\rm{M}^2_{\rm{W}})$=$0.119\pm0.002$, the measured
leptonic branching fraction of the W yields
\begin{equation}
  \sum_{\tiny\begin{array}{c}i=(u,c),\\j=(d,s,b)\\\end{array}}
  |\rm{V}_{ij}|^2
  \quad = \quad
  2.026\,\pm\,0.026\pm\,0.001,
\label{res}
\end{equation}
where the first error is due to the uncertainty on the branching
fraction measurement and the second to the uncertainty on
$\alpha_s$ \cite{lep}. This result is consistent with the
unitarity of the first two rows of the CKM matrix at the 1.5\%
level, as in the Standard Model:
\begin{equation}
          \sum_{\tiny\begin{array}{c}i=(u,c),\\j=(d,s,b)\\\end{array}}
          |\rm{V}_{ij}|^2 = 2 .
\label{th}
\end{equation}
 No assumption on the values of the single CKM elements are made.

Using the experimental knowledge~\cite{Hagiwara} of the sum
$|\rm{V}_{ud}|^2+|\rm{V}_{us}|^2+|\rm{V}_{ub}|^2+
 |\rm{V}_{cd}|^2+|\rm{V}_{cb}|^2=1.0477\pm0.0074$,
the above result can also be interpreted as a measurement of
$|\rm{V}_{cs}|$:
$$
  |\rm{V}_{cs}|\quad=\quad0.989\,\pm\,0.014.
$$
The error includes a $\pm0.0006$ contribution from the uncertainty
on $\alpha_s$ and a $\pm0.004$ contribution from the uncertainties
on the other CKM matrix elements, the largest of which is that on
$|\rm{V}_{cd}|$. These contributions are negligible compared to
the experimental error from the measurement of the W branching
fractions, $\pm0.014$.


The  W decay branching fractions,
\mbox{$\mathcal{B}(\rm{W}\rightarrow\textrm{f}\overline{\textrm{f}}')$},
are determined from the cross sections for the individual
WW$\rightarrow\rm{4f}$ decay channels measured by the four
experiments at all energies above 161 GeV.
These branching fractions can be derived with and without the
assumption of lepton universality. In the fit with lepton
universality, the branching fraction to hadrons is determined from
that to leptons by constraining the sum to unity.



Assuming lepton universality, the measured hadronic branching
fraction is $67.77\pm0.18\rm{(stat.)}\pm0.22\rm{(syst.)}\%$ and
the measured leptonic branching fraction is
$10.74\pm0.06\rm{(stat.)}\pm0.07\rm{(syst.)}\%$. These results are
consistent with the Standard Model expectations, 67.51\% and
10.83\% respectively~\cite{yellowreport}. In this case, the high
$\chi^2$ of 20.8 for 11 degrees degrees of freedom is mainly due
to the \Ltre\ results for W decays to muons and taus.

\section{Conclusion}
The presently poorly satisfied unitarity condition for the CKM
matrix presents a puzzle in which a deviation $\Delta$ from
unitarity may point towards new physics. The unitarity tests fails
by up to 2.7 standard deviations. The origin of deviation $\Delta$
is unclear.

In $|V_{ud}|$ from nuclear $0^{+} \rightarrow 0^{+}$ transitions
with deviation $\Delta$ = 0.0032$\pm$0.0014, the error is
dominated by theoretical uncertainties, where errors of nuclear
structure dependent corrections are no larger than those of
transition independent corrections. Restoration of unitarity would
require a 2.3 $\sigma$ shift of these corrections. In $|V_{ud}|$
from neutron decay, with deviation $\Delta$ = 0.0076$\pm$0.0028,
the error is dominated by experimental uncertainties. Restoration
of unitarity would require a 3 $\sigma$ shift in the present value
of the $\beta$-decay asymmetry $A$, or a 8 $\sigma$ shift in the
neutron lifetime $\tau$. Alternatively, radiative corrections
would have to be wrong by 8 $\sigma$. If the deviation is due to
errors in $|V_{us}|$, its presently accepted value would have to
shift by 7 $\sigma$ in order to explain the neutron result, or by
3 $\sigma$ to explain the nuclear $0^{+} \rightarrow 0^{+}$
result. However, very recent preliminary results hint that the
last world on $|V_{us}|$ is not yet spoken. On the other hand,
$|V_{ub}|$ is completely negligible in this context. Other sources
such as pion or W decay may have smaller theoretical errors, but
their present experimental errors are not yet competitive.

\addcontentsline{toc}{section}{References}

\section{Acknowledgements} The work on nuclear $\beta$-decay was
done in collaboration with I.S. Towner. It was supported by the
U.S. Department of Energy under Grant No. DE-FG03-93ER40773 and by
the Robert A. Welch Foundation. J.C.H. should also like to thank
the Institute for Nuclear Theory at the University of Washington
for its hospitality and support during part of this work. The work
of W.J.M. was supported by DOE contract DE-AC02-98-CH10886 and a
Senior Research Award from the Humboldt Foundation. The work on
neutron $\beta$-decay was funded by the German Federal Ministry
for Research and Education under contract number 06HD953. The two
day workshop \emph{\scshape{Quark-Mixing, CKM-Unitarity}} was
sponsored by the Deutsche Forschungsgemeinschaft (German Research
Foundation).


\begin{thebibliography}{}
\bibitem{Soldner}T. Soldner, Recent progress in neutron polarization and its
analysis, ref. \cite{HADM}
\end{thebibliography}
\end{document}